\documentclass[%
reprint,
groupedaddress,
openany
bibnotes,
 amsmath,amssymb,
 aps,
prl,
]{revtex4-1}

\usepackage{graphicx}
\usepackage{epstopdf}
\usepackage{dcolumn}
\usepackage{bm}

\begin{document}

\preprint{APS/123-QED}

\title{Variation of Tensor Force due to Nuclear Medium Effect}
\email{tfwang@buaa.edu.cn (T. F. Wang)}
 \author{Ziming Li$^{1}$, Jie Zhu$^{1}$, Taofeng Wang$^{1*}$,  Minliang Liu$^{2}$, Jiansong Wang$^{2}$,\\ Yanyun Yang$^{2}$, Chengjian Lin$^{3}$, Zhiyu Sun$^{2}$, Qinghua He$^{4}$, M. Assié$^{5}$,Y. Ayyad$^{6}$,\\ D. Beaumel$^{5}$, Zhen Bai$^{2}$, Fangfang Duan$^{2}$, Zhihao Gao$^{2}$, Song Guo$^{2}$, Yue Hu$^{1}$,\\ Wei Jiang$^{7}$,  F. Kobayashi$^{8}$, Chengui Lu$^{2}$, Junbing Ma$^{2}$, Peng Ma$^{2}$,\\ P. Napolitani$^{9}$, G. Verde$^{10,11}$, Jianguo Wang$^{2}$, Xianglun Wei$^{2}$, Guoqing Xiao$^{2}$,\\ Hushan Xu$^{2}$, Biao Yang$^{7}$, Runhe Yang$^{2}$, Yongjin Yao$^{1}$, Chaoyue Yu$^{2}$,\\ Junwei Zhang$^{2}$, Xing Zhang$^{2}$, Yuhu Zhang$^{2}$,  Xiaohong Zhou$^{2}$}
\affiliation{%
 $^{1}$School of Physics, Beihang University, Beijing 100191, China\\ 
 $^{2}$Institute of Modern Physics, Chinese Academy of Sciences, Lanzhou 730000, China\\
 $^{3}$China Institute of Atomic Energy, P.O. Box 275 (10), Beijing 102413, China\\
 $^{4}$Department of Nuclear Science $\&$ Engineering, College of Material Science and Technology,\\Nanjing University of Aeronautics and Astronautics, Nanjing 210016, China\\ 
 $^{5}$IJCLab, Université Paris-Saclay, CNRS/IN2P3, 91405 Orsay, France\\
 $^{6}$Facility for Rare Isotope Beams, Michigan State University, East Lansing, Michigan 48824, USA\\
 $^{7}$State Key Laboratory of Nuclear Physics and Technology, School of Physics, Peking University, Beijing 100871, China\\
 $^{8}$Graduate School of Engineering Science, Osaka University, 1-3 Machikaneyama, Toyonaka, Osaka 560-8531, Japan\\
 $^{9}$IPN, CNRS/IN2P3, Université Paris-Sud 11, Université Paris-Saclay, 91406 Orsay Cedex, France\\
 $^{10}$INFN Sezione di Catania, via Santa Sofia 64, I-95123 Catania, Italy\\
 $^{11}$Laboratoire des 2 Infinis - Toulouse (L2IT-IN2P3), Université de Toulouse, CNRS, UPS, F-31062 Toulouse Cedex 9, France
  }

\begin{abstract}
The enhancement of $J^{\pi}(T)$=3$^{+}$(0) state with isospin $T=0$ excited by the tensor force in the free $^{6}$Li nucleus has been observed, for the first time, relative to a shrinkable excitation in the $^{6}$Li cluster component inside its host nucleus. Comparatively, the excitation of $J^{\pi}(T)$=0$^{+}$(1) state with isospin $T=1$ for these two $^{6}$Li formations take on an approximately equal excitation strength. The mechanism of such tensor force effect was proposed due to the intensive nuclear medium role on isospin $T$=0 state.


\end{abstract}

\pacs{Valid PACS appear here}
\maketitle


Tensor force has been proved to be of importance in reproducing the properties of nuclear matter and explaining the D-wave mixing and binding energy in deuterons. Some studies have shown that the tensor force may play a significant role in the properties of the new magic number of neutron rich nuclei and the order changes of single particle orbits. Tensor force essentially make variations in single-particle energy causing the shell evolution [1]. Tensor force also contributes to the reduction of the spin-orbits splitting [2, 3] and to the attractive role in the nuclear binding energy [4].

The studies of the short-range correlation indicate that the high momentum component in the wave functions of the nucleon momentum distribution larger than the Fermi momentum mainly comes from the tensor force inductions of proton-neutron correlated pairs [5, 6]. The spins of neutron and proton ($\boldsymbol {s_{n}}$, $\boldsymbol {s_{p}}$) can be either parallel ($I$=1; $l$=0, 1, 2) or antiparallel ($I$=0; $l$=1), where $\boldsymbol {I=s_{n}+s_{p}}$ and $\boldsymbol {J=I+l}$. There are four ways to couple $\boldsymbol {s_{n}}$, $\boldsymbol {s_{p}}$ and $\boldsymbol {l}$ to get a measured total $J$ = 1 for deuteron. The parity of deuteron is associated with the orbital motion of (-1)$^{l}$. The observed even parity for deuteron leads to eliminate the combination of spins that include $l$=1, leaving $l$=0 and $l$=2 as possibilities. The wave function of the deuteron is therefore consist of the mixture of S and D components. 

The spin and isospin of nucleons in proton-neutron ($pn$) pair can be combined into different channels. Tensor force has a strong population strength for the spin and isospin for $J, T$ = 1, 0 channel, but a weak one for $J, T$ = 0, 1 channel [7]. The proton-neutron tensor monopole interaction ($T$ = 0) is twice as strong as the $T$ = 1 interaction [8]. For the same radial condition of wave function, larger orbital angular momentums of proton and neutron subshell may intensively enhance the tensor monopole effect when their relative momentum becomes higher [1, 8].

The interaction properties of the tensor force component depend on the orientation direction of spins of proton and neutron relative to the direction of the vector connecting them [9]. The tensor force conducts between two nucleons with their spins aligned, and the interaction is attractive when the spins are parallel to the line connecting the two nucleons and repulsive when the spins are perpendicular to this line [9].

The configuration of the two-body cluster in $^{6}$Li can be represented by p-shell nucleons of $pn$ pair (deuteron)  coupled to s-shell nucleons of the $\alpha$ core  with intrinsic spin 1 and intrinsic orbital angular momentum $l$ = 0 as well as relative momentum $L$ = 0, generating the 1$^{+}$ ground state and the excited triplet states 1$^{+}$, 2$^{+}$, 3$^{+}$ with isospin $T$ = 0 of $^{6}$Li from $L$ = 2 [10]. The cluster-cluster spin-orbit force splits the $L$ = 2 as these triplet states, where $L$ the orbital momentum coupling of $pn$ pair and $\alpha$. The small magnitude with a negative sign for the electric quadrupole moment of $^{6}$Li was suggested to be accounted for by introducing the cluster-cluster tensor interaction which mixes a small $L$ = 2 component into the $^{6}$Li states [10].

$^{11}$C is a proton-rich isotope with six protons and five neutrons. This asymmetry in the proton-to-neutron ratio can significantly affect the stability and behavior of the nucleus. The strong Coulomb repulsion between protons makes this nuclei prone to form the cluster structure with the general formalism of $\alpha$ and the tensor force favored $pn$ pair of $d$. $d$+$\alpha$ are feasible to form the weakly bound $^{6}$Li nuclei within $^{11}$C.

In this letter, we present an inspired aspect of tensor force effect enhancement drived by isospin $T$ = 0 state with large spin $J^{\pi}$ = 3$^{+}$ in the free $^{6}$Li, while a shringking effect evidently in the $^{6}$Li cluster inside nucleus due to the nuclear medium. Comparatively, the isospin $T$=1 with $J^{\pi}$ = 0$^{+}$ state maintains an equal excitation strength in these two different $^{6}$Li formations.

The present experimental measurement was performed at the Radioactive Ion Beam Line at the Heavy Ion Research Facility in Lanzhou (HIRFL-RIBLL) [11]. A 60 MeV/nucleon $^{12}$C beam was transfed to bombard a 3.5 mm $^{9}$Be target to produce about 25 MeV/nucleon $^{11}$C secondary beam with a purity of about 99$\%$ and an intensity of about 10$^{4}$ particles per second. The beam particles were identified in terms of $B\rho-$TOF$-\Delta E$ method with the magnets and two plastic scintillator detectors in the beam line [26]. The $^{11}$C secondary beam were bombarded on a 50 mg/cm$^{2}$ carbon target to produce the breakup reaction. 

Three parallel plate avalanche chambers (PPACs) with 50$\times$50 mm$^{2}$ active area and position resolutions of about 1 mm (FWHM) in both the $X$ and $Y$ directions were placed in front of the target to track the incident $^{11}$C beam [12] and to subsequently get the reaction vertex in the target. $d$ and $\alpha$ are detected by the zero-degree telescope system which consists of a double-sided silicon strip detector (DSSD, of 148 $\mu$m in thickness and 50$\times$50 mm$^{2}$ in cross-sectional area) with 32 strips on both front and back sides, and a 2$\times$2 photodiode (PD) readout CsI (Tl) scintillator (25$\times$25$\times$30 mm$^{3}$ size for each unit) array. Each CsI (Tl) scintillator is covered by two layers of high reflection Tyvek papers and a 10 $\mu$m aluminum coated Mylar film as window. The PD is coupled to CsI (Tl) scintillator with the photoconductive silicone grease. The angular coverage of the zero-degree telescope is about 0-9$^{\circ}$. Five LaBr$_{3}$ (Ce) and one NaI scintillator detectors were placed around the target to measure the decayed $\gamma$s from the excited fragments. DSSD was utilized to record the $\Delta$E energy and the position of the detecting fragments, therefore, the emission angle may be obtained by combining with the reaction vertex in the target. CsI (Tl) detection system provides the residual E energy of the fragments. Particle identifications (PID) for $\alpha$ and $d$ were performed using $\Delta$E-E contour. The energy resolution with sigma of this  $\Delta$E-E detection system is estimated to be $\sim$0.8 MeV from numerical simulation.

\begin{figure}[htb]
\includegraphics[width=9.0cm]{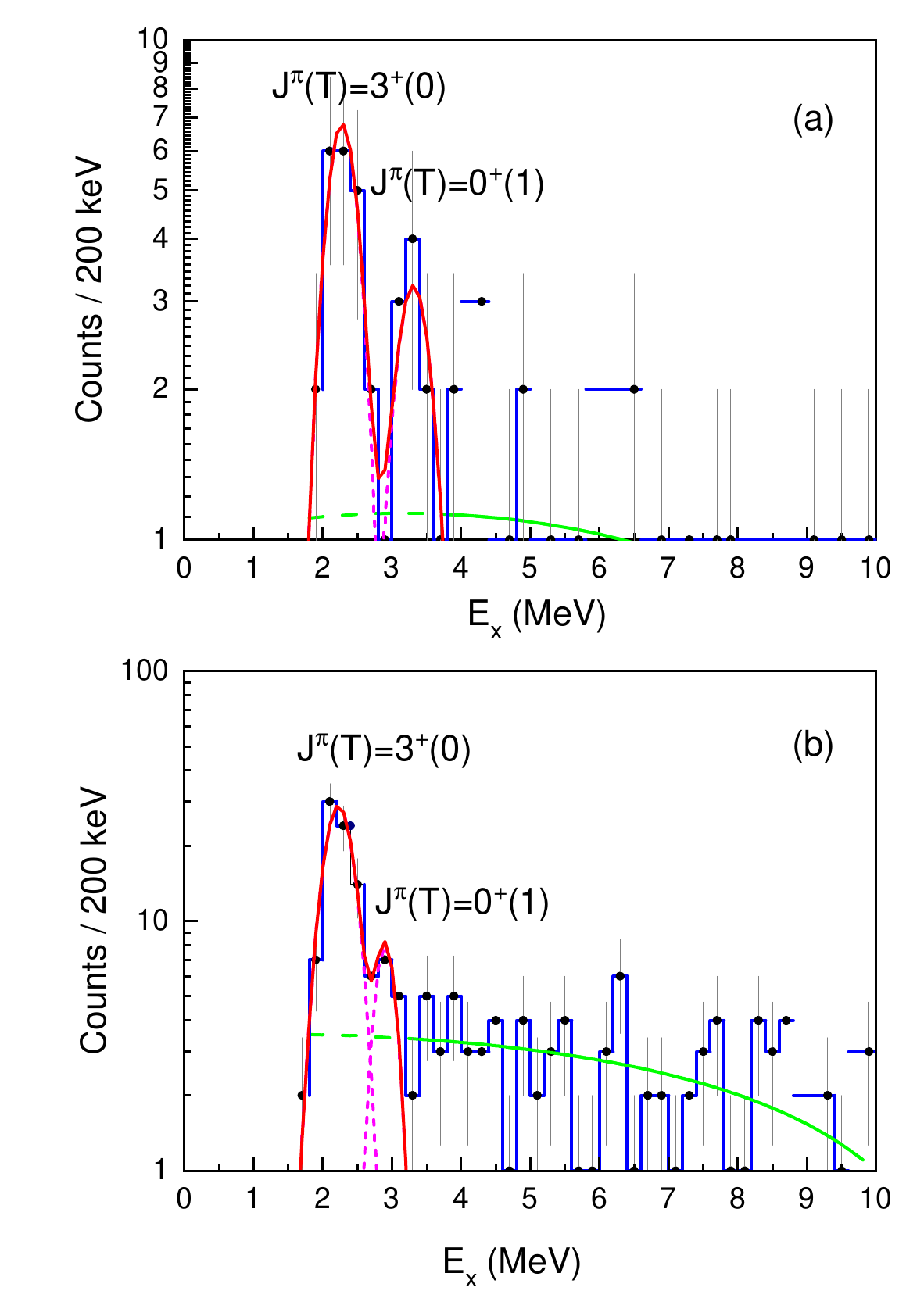}
\caption{\label{fig:pdfart}  Excitation energy spectra of $^{6}$Li$^*$: (a) Under the cut of $\Theta_{CM}<120^{\circ}$ where $\Theta_{CM}$ is the relative emitting angle between $d$ and $\alpha$ in the center-of-mass-system of $^{6}$Li, clear excited states of $J^{\pi}(T)=3^{+}(0)$ and $J^{\pi}(T)=0^{+}(1)$ are observed;  (b) Under the cut of $\Theta_{CM}>120^{\circ}$ there exhibits a enhanced population of $J^{\pi}(T)=3^{+}(0)$ state, while $J^{\pi}(T)=0^{+}(1)$ state keeps a similar population to that in (a). The purple dash lines are the fitting for these resonances with Breit-Wigner function, the continuous background is assigned by the green line.}
\end{figure}

\begin{figure}[htb]
\includegraphics[width=9.2cm]{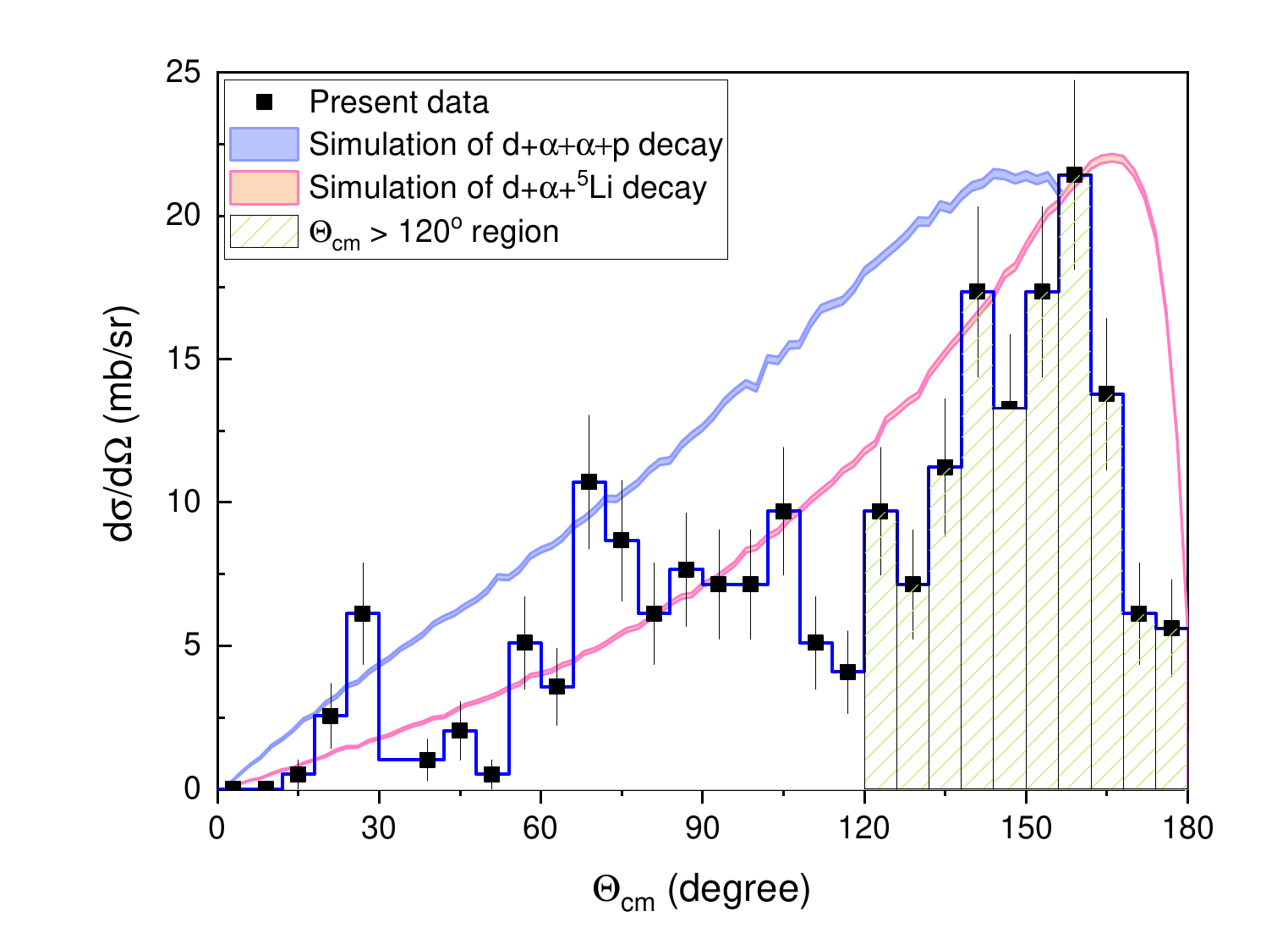}
\caption{\label{fig:pdfart} The relative emitting angle $\Theta_{CM}$ between $d$ and $\alpha$ in the center-of-mass-system of $^{6}$Li. There exist two regions separated by $120^{\circ}$. The smaller $\Theta_{CM}$ region corresponds to $d$ and $\alpha$ emitted by the $^{6}$Li cluster component initially inside $^{11}$C, comparatively, the larger $\Theta_{CM}$ reflects the back-to-back emission of $d$ and $\alpha$ from the flying $^{6}$Li isolated from $^{11}$C. }
\end{figure}


The excited state of $^{6}$Li$^*$ was reconstructed by the invariant mass of final state particles of $d$ and $\alpha$ selected with the multiplicity-2 hits events recorded by the zero-degree telescope system. The excitation energy spectra of $^{6}$Li$^*$ are investigated with two cases (Fig. 1) in terms of the relative emitting angle $\Theta_{CM}$ between $d$ and $\alpha$ in the center-of-mass-system of $^{6}$Li: (a) $\Theta_{CM}<120^{\circ}$; (b) $\Theta_{CM}>120^{\circ}$. 

Under the case of $\Theta_{CM}<120^{\circ}$, the resonance states of $J^{\pi}(T)=0^{+}(1)$ and $3^{+}(0)$ can be clearly observed, and their magnitude are comparable (Fig. 1 (a)).  Contrastingly,  the cross section of $J^{\pi}(T)=3^{+}(0)$ under the case of $\Theta_{CM}>120^{\circ}$ (Fig. 1 (b)) extremely enhances relative to that under the case of $\Theta_{CM}<120^{\circ}$ (Fig. 1 (a)). While the magnitude of the cross sections of $J^{\pi}(T)=0^{+}(1)$ state are similar between these two cases. The small relative angle $\Theta_{CM}<120^{\circ}$ indicates $d$ and $\alpha$ arise from $^{11}$C direct breakup process, since their momenta should be balanced by other fragments. In the case of $\Theta_{CM}>120^{\circ}$, the large relative emission angle implies that $d$ and $\alpha$ are emitted by back-to-back emission mode and $d$ and $\alpha$ are decay products from the flying $^{6}$Li nucleus isolated from $^{11}$C. The flying isolated $^{6}$Li$^{\ast}$ was confirmed by the clear asymmetry edge of the decay $\gamma$ spectrum of $^{6}$Li$^{\ast}$ due to the Doppler-shift effect.

The cross section ratio between $J^{\pi}(T)=3^{+}(0)$ state and $J^{\pi}(T)=0^{+}(1)$ state varies from 2.7 at $\Theta_{CM}<120^{\circ}$ to 5.5 at $\Theta_{CM}>120^{\circ}$, implying a remarkable property: the isospin $T$=0 state of $J^{\pi}=3^{+}$ in the isolated $^{6}$Li is extremely strengthened, however, this state is hindered extremely in the $^{6}$Li cluster component inside $^{11}$C. Comparatively, the isospin $T$=1 state of $J^{\pi}=0^{+}$ keeps an almost equal excitation under these two different $^{6}$Li formations. 

The transverse momentum $p_{T}$ of $d$ or $\alpha$ fragments reflects the dynamical essentiallities of clusters in the initial state in the $^{6}$Li. $p_{T}$ is also equal to the transfer momentum $\Delta p$ in the $^{6}$Li breakup reaction, and therefore,

\begin{equation}
\begin{split}
sin\Theta_{CM}=\frac{\Delta p}{p}=\frac{F\Delta t}{p}\simeq\frac{1}{p}\frac{V_{0}}{R}\frac{R}{v}=\frac{V_{0}}{pv}=\frac{V_{0}}{2E_{k}},
\end{split}
\end{equation}

where $p$ is the momentum of the $d$ or $\alpha$ fragment. $F$ the average force between $d$ and $\alpha$ is proportional to $\frac{V_{0}}{R}$, where $V_{0}$ and $R$ are the depth and range of potential between $d$ and $\alpha$. Assume the velocity of the $d$ or $\alpha$ cluster is $v$, the lifetime of the $^{6}$Li excited state can be expressed as $\tau$ = $\Delta t=\frac{R}{v}$. $E_{k}$ is the kinematic energy of $d$ or $\alpha$. According to Equ. (1), the $J^{\pi}(T)$=$3^{+}(0)$ excited state with the open angle of $d$ and $\alpha$ of $\Theta_{CM}<120^{\circ}$ corresponding to $^{6}$Li inside $^{11}$C have a longer lifetime than that of $\Theta_{CM}>120^{\circ}$ corresponding to the isolate $^{6}$Li. It is also reasonable to deduce that $d$ or $\alpha$ acts a slower dynamics with a smaller $E_{k}$ in the $^{6}$Li component in the nuclear medium relative to in the isolate $^{6}$Li.

\begin{figure}[htb]
\includegraphics[width=9.2cm]{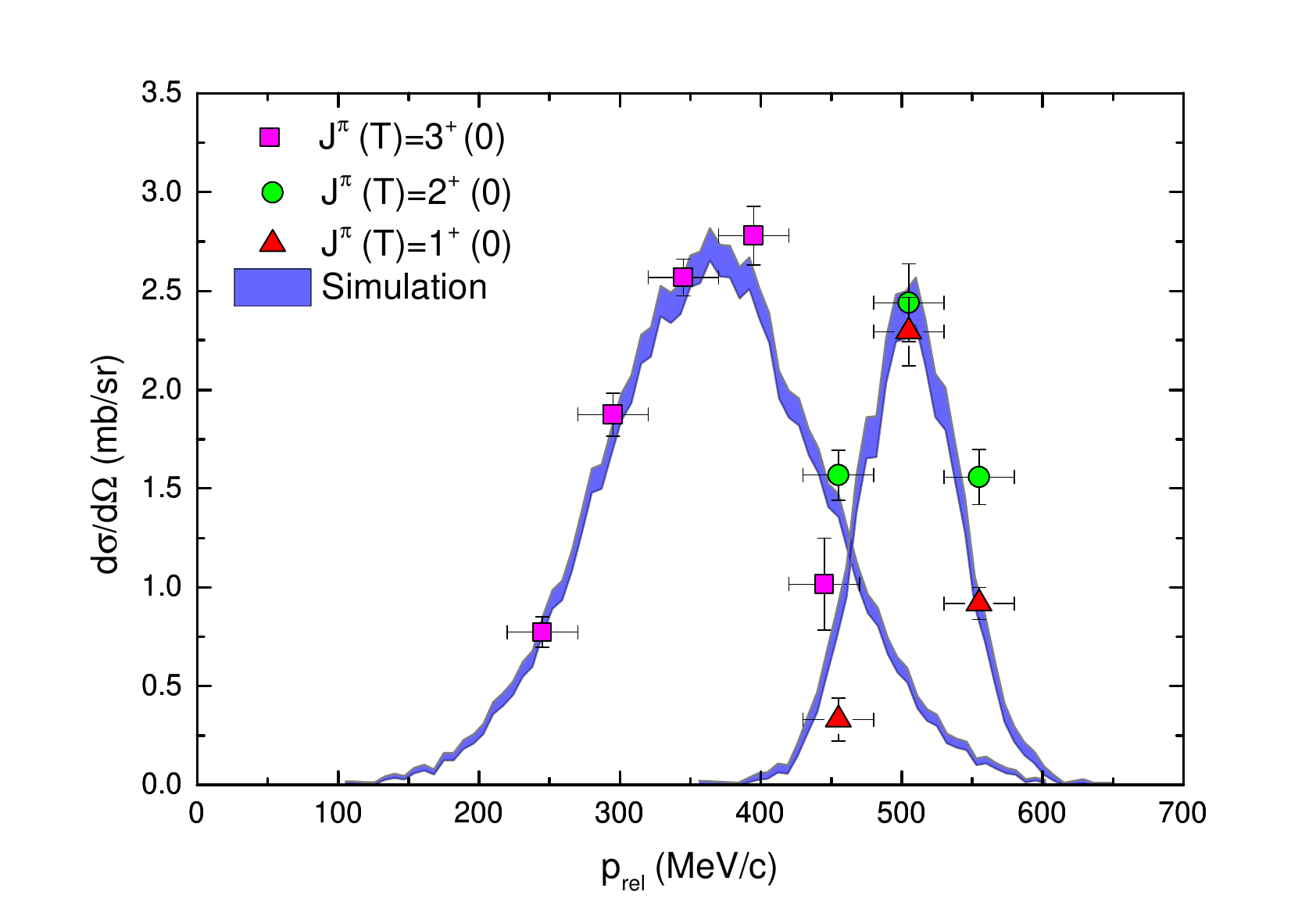}
\caption{\label{fig:pdfart} The relative momentum $p_{rel}$ between $d$ and $\alpha$ dependent on the spin of the triplet state with $T$=0. $J^{\pi}(T)$=3$^{+}(0)$ state lies in the lower $p_{rel}$ region with a wider width, $J^{\pi}(T)$=2$^{+}(0), 1^{+}(0)$ mixing states locate in the higher $p_{rel}$ region with a narrower width.}
\end{figure}

Nucleons correlation and clustering are the universal characteristic in nuclei, especially the existing molecular-like states in the neutron-rich light nuclei. The cluster indicates a spatially localized subsystem consisted of strongly correlated nucleons, such as Deutron-like short range correlated pair and the compact four-nucleon correlated $\alpha$ cluster. The relative motion between clusters become as a remarkable quantity for the fundamental motion mode of the nucleus.

The cluster model is treated based on a different degree of freedom than on the other models, typically the one-center mean field shell model. The microscopic cluster keep a dynamical process in that the everlasting change structure due to its formation, growth and breakup complemented with the shell-model-like states simultaneously. It is a significant fact that the description for such transition of structure is identified in a finite quantum many-body system of the atomic nucleus. The inter-cluster relative motion and its coupling with the excitation modes of clusters correspond to not only the bound-state but also to the highly excited states like molecular resonances.

As shown in Fig. 3, the distribution of relative momentum $\vec p_{rel}=\vec p_{\alpha}-\vec p_{d}$ between $d$ and $\alpha$ in the center-of-mass system of $^{6}$Li at $3^{+}(0)$ state lies in the low $p_{rel}$ region, while $2^{+}(0)$ and $1^{+}(0)$ mixing states locate at high $p_{rel}$ region. With Gaussian fitting, the mean value of $3^{+}(0)$ state was given to be 361.6 $\pm$ 12.7 MeV/c, and the standard deviation $\sigma$ is 77.2 $\pm$ 5.5 MeV/c. And the mean value of the mixing state of $2^{+}(0)$ and $1^{+}(0)$ is 517.7 $\pm$ 9.1 MeV/c, and $\sigma$ of 35.5 $\pm$ 2.8 MeV/c. It indicates that the low relative momentum for $d$ and $\alpha$ with a large width of momentum distribution in $3^{+}(0)$ state correspond to a weak relatively dynamical motion with a narrow pn pair spatial distribution; the higher relative momentum with a smaller width for $2^{+}(0)$ and $1^{+}(0)$ mixing states maintain a drastic relative motion and a broad occupation for the pn pair + $\alpha$ cluster in $^{6}$Li.


The tensor force strength for $J^{\pi}(T)=0^{+}(1)$ state is much smaller than that for $J^{\pi}(T)=3^{+}(0)$ state in the isolated $^{6}$Li [13], which can be essentially explained by the tensor force properties of the isospin dependence [7]. For the $0^{+}(1)$ state, which has a configuration of $(0s_{1/2})^{4}(0p_{3/2})^{2}$, the inner $^{4}$He core composed of two $pn$ pairs with $T$=0 is excited from 0$s$ to 0$p_{1/2}$ by the tensor force via two-particle two-hole (2p2h) mode, while the outer $pn$ pair on the 0$p$ shell with $T$=1 takes a relatively small excitation due to the weak tensor force for the $T$=1 channel. For the $3^{+}(0)$ state, the outer $pn$ pair in the 0$p$ shell may take place a tensor force excitation with a $T$=0 state from $(0p_{3/2})^{2}$ to $(0d_{3/2})^{2}$, which satisfy the selection rule for the tensor force excitation of $\Delta L$=2, $\Delta S$=2 and $\Delta J$=0. From the point view of excitation strength of tensor force, the coupling of outer $pn$ pair to the inner $^{4}$He core is relatively weak for $=0^{+}(1)$ state and hence the population of $3^{+}(0)$ state is much higher than that of $0^{+}(1)$ state in the isolated $^{6}$Li nucleus, as shown in Fig. 2(b).

However, when $^{6}$Li is not completely isolated but still within the domain of  $^{11}$C as a component, $d$ and $\alpha$ in $^{6}$Li are correlated two-body clusters and essentially affected by the central interactions from the residual nucleus. Such extra interactions may reduce the strength of coupling of the $pn$ pair with isospin $T$ = 0 to $\alpha$, which leads to the shrinkage of tensor force excitation and results in a comparable populations between $J^{\pi}(T)=0^{+}(1)$ state and $J^{\pi}(T)=3^{+}(0)$ state for the case of $\Theta_{CM}<120^{\circ}$, as shown in Fig. 1(a).

Among the $T$=0 triplet states of $J^{\pi}(T)=3^{+}(0),2^{+}(0),1^{+}(0)$, the strong excitation of $3^{+}(0)$ state with the lowest excitation energy results from tensor force excitation  from $(0p_{3/2})^{2}$ to $(0d_{3/2})^{2}$ with 2p2h excitation mode at a high spin and $T$=0. The highest excited energy state $1^{+}(0)$ involves the mixing configurations of $(0s_{1/2})^{4}(0p_{3/2})^{2}$, $(0s_{1/2})^{4}(0p_{1/2})^{1}(0p_{3/2})^{1}$ and $(0s_{1/2})^{4}(0p_{1/2})^{2}$ with a similar probability each other, and $2^{+}(0)$ state only takes on a configuration of $(0s_{1/2})^{4}(0p_{1/2})^{1}(0p_{3/2})^{1}$ with two valence nucleons of neutron and proton lying in the separated sub-shell levels [13], which could explain the present experimental observations of the high population of $3^{+}(0)$ state and the low population of $2^{+}(0)$ and $1^{+}(0)$ states.  In addition, This could also clarify the relatively large dynamical motion and broader density distribution of $d$ and $\alpha$, namely, a more loosely bound structure for $2^{+}(0)$ and $1^{+}(0)$ states in $^{6}$Li.

The lightest even neutron rich nucleus is considered to be the best nuclear molecule object. The valence neutrons may occupy the $\sigma$- or $\pi$- valence orbits around the separated $\alpha$ clusters. The $\pi$- valence neutrons move in a circular orbit centered on the deformation axis, while the density of the motion orbit of the $\sigma$- valence neutrons increases along the axis between the $\alpha$s. Because of Pauli's principle, the neutrons in the $\sigma$ orbit tend to push the $\alpha$ cluster far away, which leads to the more deformed structure. As concerned to $^{6}$Li, it is the lightest one-centered $\alpha$ cluster with a valence pn pair, the $\alpha$ core repulsively deviates away from the the $\sigma$-orbital valence neutron. The formation of $^{6}$Li is really an oblate shape which is reflected by the very small magnitude with a negative sign for electric quadrupole moment of $^{6}$Li [14-16].

\begin{figure}[htb]
\includegraphics[width=9.2cm]{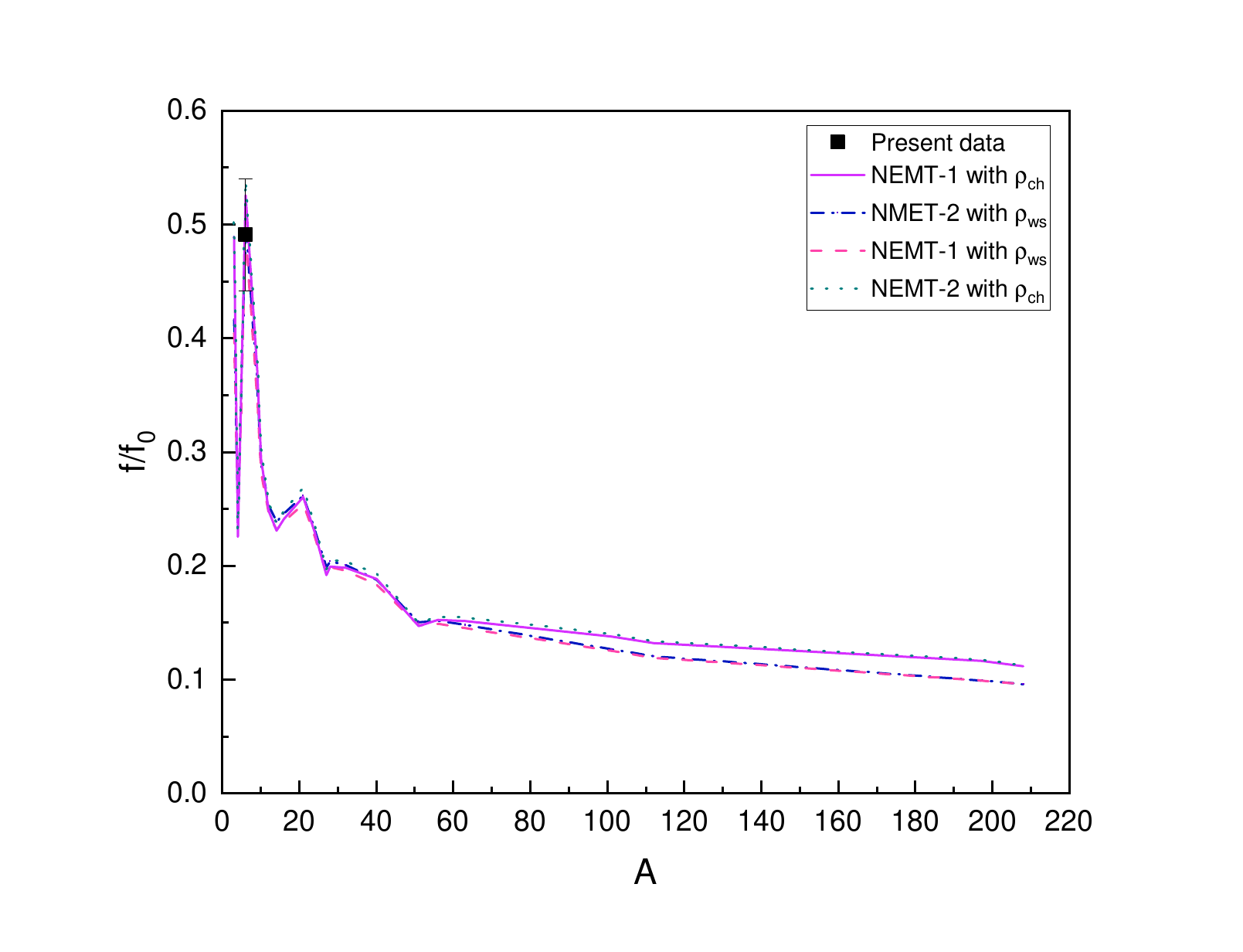}
\caption{\label{fig:pdfart} The tensor force strength ratio of without the nuclear medium to within the nuclear medium $f/f_{0}$ depending on the atomic number A, where NEMT model calculations are performed with two functions of nuclear density $\rho_{ch}$ and $\rho_{ws}$. }
\end{figure}

The tensor force strength, modulated by the nuclear medium effects, reflects the intricate interplay between the nucleons within the nucleus. As the nuclear density increases, more nucleons are packed into a given volume, leading to increased overlap and interaction between them. This enhanced interaction modifies the strength of the tensor force and its impact on the nuclear system.

\begin{equation}
\begin{split}
\frac{f(\rho)}{f_{0}(\rho)}=e^{-(a\rho+b\rho^{2}+c\rho^{3})}
\end{split}
\end{equation}

The function $f(\rho)/f_{0}(\rho)$ of nuclear medium effect for tensor force strength (NMET) in Equ. (2) captures this behavior by introducing the scaling constant $a,b,c$. The positive values of which indicate that the tensor force strength diminishes as the nuclear density increases. This decrease is a manifestation of the saturation of the strong nuclear force at high densities, where the repulsive interactions between nucleons begin to dominate, counteracting the attractive tensor force. The density distributions of the nucleus which are usually described by a Woods-Saxon type distribution as
\begin{equation}
\begin{split}
\rho_{ws}(r)=\rho_{0}\left [1+exp\left (\frac{r-R}{a}\right )\right ]^{-1}
\end{split}
\end{equation}
with diffuseness parameter $a\sim$ 0.53 fm and $\rho_{0}\sim$ 0.17 fm$^{-1}$ is the density at the center of the nucleus. The radius parameter $R$ is formed by mass number $A$ as $R\sim1.10A^{1/3}$ (fm). A charge density function was approximated as Equ. (4), which treats the shape of the nuclei as a uniform sphere with a constant charge density [17].

\begin{equation}
\begin{split}
\rho_{ch}(r)=\frac{0.75A/\pi}{(5/3<r^{2}>)^{3/2}}
\end{split}
\end{equation}

The calculations of with and without $b$, $c$ terms in Equ. (2) defining as NMET-1 and NMET-2 models have been performed with two nuclear density functions of Equ. (3) and Equ. (4), as shown in Fig. 4. The parameters were ajusted according to the scaling of tensor force with nuclear density to consistent to the present experimental data of 3$^{+}$(0) state of $^{6}$Li. A saturation trend is achieved at high densities, reflecting the intricate behavior of the strong nuclear force within atomic nuclei.    

The modification of tensor force strength with nuclear density effects becomes particularly important in extreme conditions, such as neutron stars or during heavy-ion collisions in nuclear physics experiments. In these extreme environments, the nuclear density can reach values far beyond those found in stable atomic nuclei. The modification of the tensor force strength with nuclear density effects influences the equation of state, nuclear structure, and transport properties of nuclear matter.


Summarily, a shrink excitation for spin $J^{\pi}=3^{+}$ state with isospin $T=0$ was observed in the $^{6}$Li cluster component inside the nucleus for the first time, which distinctly differs from the relatively enhanced excitation of this state in the isolated nucleus $^{6}$Li. While the excitation of the reference state of $J^{\pi}=0^{+}$ with isospin $T=1$ exhibits an almost equal cross section for these two $^{6}$Li formations. The mechanism of this relatively weaker excitation in the $^{6}$Li cluster in $^{11}$C nucleus is proposed due to the nuclear medium effects hindering the tensor force excitation for the isospin $T$ = 0 state, but no impact on the $T=1$ state. 

We would like to acknowledge the staff of HIRFL for the operation of the cyclotron. The author T. Wang appreciates for the financial supports from China Scholarship Council. This work has also been supported by the National Natural Science Foundation of China (No. 10175091 and No. 11305007).


\end{document}